\documentclass[twocolumn,showpacs,preprintnumbers,amsmath,amssymb,eqsecnum]{revtex4}
\usepackage{dcolumn}
\usepackage{bm}
\usepackage[dvips]{graphicx}
\usepackage{wrapfig}
\usepackage{psfrag}
\begin{document}

\def\sh{\mathop{\rm sh}\nolimits}
\def\ch{\mathop{\rm ch}\nolimits}
\def\ctg{\mathop{\rm ctg}\nolimits}
\def\var{\mathop{\rm var}}
\def\exp{\mathop{\rm exp}\nolimits}
\def\Re{\mathop{\rm Re}\nolimits}
\def\Sp{\mathop{\rm Sp}\nolimits}
\def\kp{\mathop{\text{\ae}}\nolimits}
\def\bk{{\bf {k}}}
\def\bp{{\bf {p}}}
\def\bq{{\bf {q}}}
\def\lra{\mathop{\longrightarrow}}
\def\Const{\mathop{\rm Const}\nolimits}
\def\sh{\mathop{\rm sh}\nolimits}
\def\ch{\mathop{\rm ch}\nolimits}
\def\var{\mathop{\rm var}}
\def\mK{\mathop{{\mathfrak {K}}}\nolimits}
\def\mR{\mathop{{\mathfrak {R}}}\nolimits}
\def\mv{\mathop{{\mathfrak {v}}}\nolimits}
\def\mV{\mathop{{\mathfrak {V}}}\nolimits}
\def\mD{\mathop{{\mathfrak {D}}}\nolimits}
\def\mN{\mathop{{\mathfrak {N}}}\nolimits}
\def\mS{\mathop{{\mathfrak {S}}}\nolimits}

\newcommand\ve[1]{{\mathbf{#1}}}

\def\Re{\mbox {Re}}
\newcommand{\Z}{\mathbb{Z}}
\newcommand{\R}{\mathbb{R}}
\def\mK{\mathop{{\mathfrak {K}}}\nolimits}
\def\mR{\mathop{{\mathfrak {R}}}\nolimits}
\def\mv{\mathop{{\mathfrak {v}}}\nolimits}
\def\mV{\mathop{{\mathfrak {V}}}\nolimits}
\def\mD{\mathop{{\mathfrak {D}}}\nolimits}
\def\mN{\mathop{{\mathfrak {N}}}\nolimits}
\def\ml{\mathop{{\mathfrak {l}}}\nolimits}
\def\mf{\mathop{{\mathfrak {f}}}\nolimits}
\def\mQ{\mathop{{\mathfrak {Q}}}\nolimits}
\def\mP{\mathop{{\mathfrak {P}}}\nolimits}
\newcommand{\ccm}{{\cal M}}
\newcommand{\cE}{{\cal E}}
\newcommand{\cV}{{\cal V}}
\newcommand{\cI}{{\cal I}}
\newcommand{\cR}{{\cal R}}
\newcommand{\cK}{{\cal K}}
\newcommand{\cH}{{\cal H}}
\newcommand{\cM}{{\cal M}}
\newcommand{\cN}{{\cal N}}

\def\br{\mathop{{\bf {r}}}\nolimits}
\def\bS{\mathop{{\bf {S}}}\nolimits}
\def\bA{\mathop{{\bf {A}}}\nolimits}
\def\bJ{\mathop{{\bf {J}}}\nolimits}
\def\bn{\mathop{{\bf {n}}}\nolimits}
\def\bg{\mathop{{\bf {g}}}\nolimits}
\def\bv{\mathop{{\bf {v}}}\nolimits}
\def\be{\mathop{{\bf {e}}}\nolimits}
\def\bp{\mathop{{\bf {p}}}\nolimits}
\def\bz{\mathop{{\bf {z}}}\nolimits}
\def\bbf{\mathop{{\bf {f}}}\nolimits}
\def\bb{\mathop{{\bf {b}}}\nolimits}
\def\ba{\mathop{{\bf {a}}}\nolimits}
\def\bx{\mathop{{\bf {x}}}\nolimits}
\def\by{\mathop{{\bf {y}}}\nolimits}
\def\br{\mathop{{\bf {r}}}\nolimits}
\def\bs{\mathop{{\bf {s}}}\nolimits}
\def\bH{\mathop{{\bf {H}}}\nolimits}
\def\bk{\mathop{{\bf {k}}}\nolimits}
\def\be{\mathop{{\bf {e}}}\nolimits}
\def\bnul{\mathop{{\bf {0}}}\nolimits}
\def\bq{{\bf {q}}}

\newcommand{\oV}{\overline{V}}
\newcommand{\vkp}{\varkappa}
\newcommand{\os}{\overline{s}}
\newcommand{\opsi}{\overline{\psi}}
\newcommand{\ov}{\overline{v}}
\newcommand{\oW}{\overline{W}}
\newcommand{\oPhi}{\overline{\Phi}}

\def\mI{\mathop{{\mathfrak {I}}}\nolimits}
\def\mA{\mathop{{\mathfrak {A}}}\nolimits}

\def\st{\mathop{\rm st}\nolimits}
\def\tr{\mathop{\rm tr}\nolimits}
\def\sign{\mathop{\rm sign}\nolimits}
\def\d{{\mathrm d}}
\def\const{\mathop{\rm const}\nolimits}
\def\O{\mathop{\rm O}\nolimits}
\def\Spin{\mathop{\rm Spin}\nolimits}
\def\exp{\mathop{\rm exp}\nolimits}
\def\mU{\mathop{{\mathfrak {U}}}\nolimits}
\newcommand{\cU}{{\cal U}}
\newcommand{\cD}{{\cal D}}
\newcommand{\cW}{{\cal W}}
\newcommand{\cQ}{{\cal Q}}
\newcommand{\cP}{{\cal P}}

\def\mI{\mathop{{\mathfrak {I}}}\nolimits}
\def\mA{\mathop{{\mathfrak {A}}}\nolimits}

\def\st{\mathop{\rm st}\nolimits}
\def\tr{\mathop{\rm tr}\nolimits}
\def\sign{\mathop{\rm sign}\nolimits}
\def\const{\mathop{\rm const}\nolimits}
\def\O{\mathop{\rm O}\nolimits}
\def\Spin{\mathop{\rm Spin}\nolimits}
\def\exp{\mathop{\rm exp}\nolimits}

\title{Wilson fermion doubling phenomenon on irregular lattice: the
similarity and difference with the case of regular lattice}

\author {S.N. Vergeles\vspace*{4mm}\footnote{{e-mail:vergeles@itp.ac.ru}}}

\affiliation{Landau Institute for Theoretical Physics,
Russian Academy of Sciences,
Chernogolovka, Moscow region, 142432 Russia \linebreak
and   \linebreak
Moscow Institute of Physics and Technology, Department
of Theoretical Physics, Dolgoprudnyj, Moskow region,
Russia}

\begin{abstract} It is shown that the Wilson fermion doubling
phenomenon on irregular lattices (simplicial complexes) does exist.
This means that the irregular (not smooth) zero or soft modes exist.
The statement is proved on 4 Dimensional lattice by means of the
Atiyah-Singer index theorem, then it is extended  easily into the
cases $D<4$. But there is a fundamental difference between doubled
quanta on regular and irregular lattices: in the latter case the
propagator decreases exponentially. This means that the doubled
quanta on irregular lattice are "bad" quasiparticles.
\end{abstract}


\maketitle

\section{Introduction}

The Wilson fermion doubling phenomenon on the regular periodic
lattices has been discovered long ago in \cite{1}. The phenomenon
and its influence on physics was studied in a number of works (for
example see \cite{2}-\cite{4}). It was proved in \cite{5}-\cite{6}
that the fermion doubling phenomenon indeed takes place on any
periodic lattice with local fermion action transforming to the usual
Dirac action in long-wavelength region. But the question about the
existence of the Wilson fermion doubling on irregular lattices is
open at present. This means that the problem is unsolved in the case
of lattice quantum gravity theory \footnote{All variants of the
lattice gravity theory are defined on the simplicial complexes} (see
\cite{7}-\cite{8}).

In this paper I show that the Wilson fermion doubling phenomenon on
irregular lattices (simplicial complexes) with $D\leq4$ does exist.
However, there exists a fundamental difference between the
propagation of doubling modes on regular and irregular lattices. In
the first case the propagator of the irregular modes is the same as
the propagator of the regular modes from the spectrum origin, i.e.
power-behaved. On the contrary, the propagation of irregular modes
on irregular lattice is similar to the Markov process of a random
walks. Thus the propagator of irregular modes on irregular lattice
decreases very quickly (exponentially): the doubled irregular modes
are "bad" quasiparticles.

\section{Fermions on irregular lattice}

First of all, one must outline shortly the Dirac system on the
simplicial complexes. More general problem (the definition of Dirac
system in discrete lattice gravity) has been solved in
\cite{7}-\cite{8}. Here I simplify the problem assuming that the
4-Dimensional simplicial complex $\mK$ is embedded into
4-Dimensional Euclidean space and the curvature and torsion are
equal to zero.

Further all definitions and designations are similar to that in
\cite{7,8}.  The four Dirac matrices $(4\times4)$ satisfy the well
known Clifford algebra
\begin{gather}
\gamma^a\gamma^b+\gamma^b\gamma^a=2\delta^{ab}, \quad
\gamma^5\equiv\gamma^1\gamma^2\gamma^3\gamma^4\,, \quad
\nonumber \\
\tr\,\gamma^5\gamma^a\gamma^b\gamma^c\gamma^d=4\,
\varepsilon^{abcd},  \quad
\sigma^{ab}\equiv\frac14\left[\gamma^a,\,\gamma^b\right].
\label{dqg80}
\end{gather}
The vertices of the complex are denoted as $a_{\cV}$, the index
${\cV}=1,2,\dots,\,{\mN}\rightarrow\infty$ enumerates the vertices.
Let the index ${\cW}$ enumerates 4-simplices. It is necessary to use
the local enumeration of the vertices $a_{\cV}$ attached to a given
4-simplex: the all five vertices of a 4-simplex with index ${\cW}$
are enumerated as
$a_{{\cW}i},\,a_{{\cW}j},\,a_{{\cW}k},\,a_{{\cW}l}$, and
$a_{{\cW}m}$, $i,\,j,\ldots=1,2,3,4,5$. It must be kept in mind that
the same vertex, 1-simplex et cetera can belong to the another
adjacent 4-simplexes. The later notations with extra index  ${\cW}$
indicate that the corresponding quantities belong to the
4-simplex with index ${\cW}$. The Levi-Civita symbol with in pairs
different indexes $\varepsilon_{{\cW}ijklm}=\pm 1$ depending on
whether the order of vertices
$a_{{\cW}i}a_{{\cW}j}a_{{\cW}k}a_{{\cW}l}a_{{\cW}m}$ defines the
positive or negative orientation of this 4-simplex.
An element of the (isotopic) gauge group
\begin{gather}
U_{{\cW}ij}=U^{-1}_{{\cW}ji}=\exp\left(ieA_{{\cW}ij}\right), \quad
A_{{\cW}ij}\in{\cal L},
\label{dqg90}
\end{gather}
where ${\cal L}$ is the Lie algebra of the gauge group, and an
elementary vector
\begin{gather}
e^a_{{\cW}ij}\equiv-e^a_{{\cW}ji},
\label{dqg100}
\end{gather}
are assigned for each
oriented 1-simplex $a_{{\cW}i}a_{{\cW}j}$. The Dirac spinors $\psi_{\cV}$ and $\psi^{\dag}_{\cV}$
are assigned to each vertex $a_{\cV}$. The Dirac spinors and the
gauge field $A_{{\cW}ij}$ belong to the same representation of
algebra ${\cal L}$.

The Euclidean Hermitean action of the Dirac field associated with
the complex $\mK$ has the form
\begin{widetext}
\begin{gather}
\mA_{\psi}=-\frac{1}{3!\,5!)}
\sum_{\cW}\sum_{i,j,k,l,m}\varepsilon_{{\cW}ijklm}\varepsilon^{abcd}
 \left(i\,\psi_{{\cW}m}^{\dag}\gamma^a
U_{{\cW}mi}\psi_{{\cW}i}\right)
e^b_{{\cW}mj}e^c_{{\cW}mk}e^d_{{\cW}ml}\equiv
\nonumber \\
\equiv\sum_{{\cV}_1{\cV}_2}\psi^{\dag}_{{\cV}_1s_1}\left[
-i\gamma^a_{s_1s_2}{\cD}^a_{{\cV}_1,{\cV}_2}
\right]\psi_{{\cV}_2s_2}\equiv\sum_{{\cV}_1{\cV}_2}\psi^{\dag}_{{\cV}_1s_1}
\left[-i\not\!\!{\cD}_{{\cV}_1,{\cV}_2}\right]_{s_1s_2}\psi_{{\cV}_2s_2}.
\label{dqg110}
\end{gather}
\end{widetext}
The indices $s_1,\,s_2=1,2,3,4$ are the Dirac one. The action
(\ref{dqg110}) is invariant under the gauge transformations
\begin{gather}
U_{{\cW}ij}\rightarrow S_{{\cW}i}\,U_{A\,ij}\,S^{-1}_{{\cW}j}, \quad
S_{{\cW}i}\in SU(2),
\nonumber \\
\psi_{{\cW}i}\rightarrow S_{{\cW}i}\,\psi_{{\cW}i}, \ \qquad \
\psi^{\dag}_{{\cW}i}\rightarrow\psi^{\dag}_{{\cW}i}\,S^{-1}_{{\cW}i}.
\label{dqg130}
\end{gather}
The curvature in (\ref{dqg110}) is equal to zero by definition. The
system of equations
\begin{gather}
\bigl(e_{ij}^a+e_{jk}^a+\ldots+e_{li}^a\bigr)=0
\label{cor350}
\end{gather}
means that the torsion is also zero. Here the sums in the
parentheses are taken on any and all closed paths. Therefore the
following interpretation is valid:
$e^a_{{\cW}ij}=\left(x^a_{{\cW}j}-x^a_{{\cW}i}\right)$, where
$x^a_{{\cW}i}$ are the cartesian coordinates of the vertex
$a_{{\cW}i}$.

Let
\begin{gather}
v_{\cW}=\frac{1}{(4!)(5!)}\varepsilon_{abcd}\varepsilon_{{\cW}ijklm}
e^a_{{\cW}mi}e^b_{{\cW}mj}e^c_{{\cW}mk}e^d_{{\cW}ml}
\label{cor354}
\end{gather}
be the oriented volume of the ${\cW}$-4-simplex and $v_{\cV}$ be the
sum of the volumes $v_{\cW}$ for that ${\cW}$-4-simplexes which
contain the vertex $a_{\cV}$. Thus the spinor space scalar product
is given by
\begin{gather}
\langle\psi_1|\psi_2\rangle=\frac15\sum_{\cV}v_{\cV}
\psi^{\dag}_{(1)\cV}\psi_{(2)\cV}.
\label{cor358}
\end{gather}
The operator $\left[i\not\!\!{\cD}_{{\cV}_1,{\cV}_2}\right]$ in
(\ref{dqg110}), as well as the operator
$\left[i\left(v_{{\cV}_1}\right)^{-1/2}\not\!\!{\cD}_{{\cV}_1,{\cV}_2}
\left(v_{{\cV}_2}\right)^{-1/2}\right]$, are Hermitian. Thus the
eigenfunction problem
\begin{gather}
\sum_{{\cV}_2}\left[i\left(\frac{1}{\sqrt{v_{{\cV}_1}}}\right)\not\!\!{\cD}_{{\cV}_1,{\cV}_2}
\left(\frac{1}{\sqrt{v_{{\cV}_2}}}\right)\right]\left(\sqrt{v_{{\cV}_2}}
\psi_{({\mP}){\cV}_2}\right)=
\nonumber \\
=\frac15\epsilon_{\mP}\left(\sqrt{v_{{\cV}_1}} \psi_{({\mP}){\cV}_1}\right)
\longleftrightarrow
\nonumber \\
\longleftrightarrow\sum_{{\cV}_2}\left[-\frac{i}{v_{{\cV}_1}}\not\!\!{\cD}_{{\cV}_1,{\cV}_2}\right]
\psi_{({\mP}){\cV}_2}=\frac15\epsilon_{\mP}\psi_{({\mP}){\cV}_1}
\label{cor363}
\end{gather}
is correct, and the set of eigenfunctions
$\left\{\psi_{({\mP})}\right\}$ forms a complete orthonormal basis
in the metric (\ref{cor358}). Let's expand the Dirac fields in this
basis:
\begin{gather}
\psi_{\cV}=\sum_{\mP}\eta_{\mP}\psi_{({\mP}){\cV}}, \quad
\psi_{\cV}^{\dag}=\sum_{\mP}\eta^{\dag}_{\mP}\psi^{\dag}_{({\mP}){\cV}}.
\label{cor366}
\end{gather}
The new dynamic variables $\{\eta_{\mP},\,\eta^{\dag}_{\mP}\}$ are
Grassmann.  The scalar product (\ref{cor358}) in these variables is
rewritten as
\begin{gather}
\langle\psi_1|\psi_2\rangle=\sum_{\mP}\eta^{\dag}_{(1){\mP}}\eta_{(2){\mP}}.
\label{cor368}
\end{gather}

It is important here that
\begin{gather}
\gamma^5i\not\!\!{\cD}_{{\cV}_1,{\cV}_2}=-i\not\!\!{\cD}_{{\cV}_1,{\cV}_2}\gamma^5.
\label{cor373}
\end{gather}

The long-wavelength limit of the theory is straightforward. To do
this one should believe the quantities $A_{{\cW}ij}$ and
$e^a_{{\cW}ij}$ as the smooth 1-forms
\begin{gather}
A_{{\cW}ij}\rightarrow A_a(x)\d x^a, \quad e^a_{{\cW}ij}\rightarrow
\d x^a
\nonumber
\end{gather}
taking the small values $A_{{\cW}ij}$ and $e^a_{{\cW}ij}$ on the
vector $e^a_{{\cW}ij}$, and substitute the smooth Dirac field
$\psi(x)$ taking the value $\psi_{\cV}$ on the vertex $a_{\cV}$ for
the set of spinors $\psi_{\cV}$. As a result the action
(\ref{dqg110}), the scalar product (\ref{cor358}) and the eigenvalue
problem (\ref{cor363}) transform to the well known expressions and
equation:
\begin{gather}
\mA_{\psi}=\int\left(-i\psi^{\dag}\gamma^a\nabla_a\psi\right)\d
x^1\wedge\d x^2\wedge\d x^3\wedge\d x^4,
\nonumber \\
\nabla_a=\partial_a+ieA_a,
\label{cor385}
\end{gather}
\begin{gather}
\langle\psi_1|\psi_2\rangle=\int \psi^{\dag}_1(x)\psi_2(x)\d^{(4)}x,
\label{cor397}
\end{gather}
\begin{gather}
-i\gamma^a\nabla_a\psi_{({\mP})}(x)=\epsilon_{\mP}\psi_{({\mP})}(x).
\label{cor400}
\end{gather}

\section{The gauge anomaly and Atiyah-Singer index theorem}

The partition function of the fermion system as the functional of
the quantities $\{e^a_{{\cW}ij}\}$ and $\{A_{{\cW}ij}\}$ is given by
integral
\begin{gather}
Z\{e^a_{{\cW}ij},\,A_{{\cW}ij}\}=\int\left(D\psi^{\dag}D\psi\right)
\exp\mA_{\psi}.
\label{d10}
\end{gather}
Here the fermion functional measure is defined according to
\begin{gather}
\left(D\psi^{\dag}D\psi\right)\equiv\prod_{\cV}
\d\psi^{\dag}_{\cV}\d\psi_{\cV}F\{e^a_{{\cW}ij}\},
\label{d20}
\end{gather}
where
\begin{gather}
d\psi_{\cV}=\prod_{\varkappa}\prod_{s=1}^4\d\psi_{{\cV}\varkappa s},
\quad d\psi^{\dag}_{\cV}=\prod_{\varkappa}
\prod_{s=1}^4\d\psi^{\dag}_{{\cV}\varkappa s}, \label{d23}
\end{gather}
and the index $\varkappa$ enumerates the components of the gauge
representation. The functional $F\{e^a_{{\cW}ij}\}$ in (\ref{d20})
can be calculated easily with the help of the metric (\ref{cor358}),
but it is not interesting here. The scalar product (\ref{cor368}) in
Grassmann variables $\{\eta_{\mP},\,\eta^{\dag}_{\mP}\}$ permits to
rewrite the measure (\ref{d20}) as below:
\begin{gather}
\left(D\psi^{\dag}D\psi\right)=\prod_{\mP}\d\eta^{\dag}_{\mP}\d\eta_{\mP}.
\label{d30}
\end{gather}

Let's study the chiral transformation of the Dirac field
\begin{gather}
\psi_{\cV}\rightarrow\exp\left(i\alpha_{\cV}\gamma^5\right)\psi_{\cV},
\quad
\psi^{\dag}_{\cV}\rightarrow\psi^{\dag}_{\cV}\exp\left(i\alpha_{\cV}\gamma^5\right).
\label{d40}
\end{gather}
Obviously, the measure (\ref{d20}) is invariant under the
transformation (\ref{d40}). Moreover, even the factors
$\left(\prod_{s=1}^4\d\psi_{{\cV}\varkappa s}\right)$ and
$\left(\prod_{s=1}^4\d\psi^{\dag}_{{\cV}\varkappa s}\right)$ of the
measure (\ref{d20}) each are invariant since the matrix $\gamma^5$
is traceless. It follows from here that the measure in right-hand
side of Eq. (\ref{d30}) is also invariant under the chiral
transformation and the corresponding Jacobian $J=1$. The last
statement permits to extract some interesting information.

Suppose the chiral transformation is infinitesimal:
$\alpha_{\cV}\rightarrow 0$. From the linearized
transformations of the Dirac field (\ref{d40}) we obtain 
linearized transformations for the variables
$\{\eta_{\mP},\,\eta^{\dag}_{\mP}\}$:
\begin{gather}
\eta_{\mP}\rightarrow\eta_{\mP}+\frac{i}{5}\sum_{\mQ}\eta_{\mQ}\sum_{\cV}\alpha_{\cV}v_{\cV}
\psi^{\dag}_{{\mP}{\cV}}\gamma^5\psi_{{\mQ}{\cV}},
\nonumber \\
\eta^{\dag}_{\mP}\rightarrow\eta^{\dag}_{\mP}+\frac{i}{5}\sum_{\mQ}
\eta^{\dag}_{\mQ}\sum_{\cV}\alpha_{\cV}v_{\cV}
\psi^{\dag}_{{\mQ}{\cV}}\gamma^5\psi_{{\mP}{\cV}}.
\label{d50}
\end{gather}
The Jacobian of this transformation is equal to
\begin{gather}
J=\left(1+\frac{2i}{5}\sum_{\cV}\alpha_{\cV}v_{\cV}\sum_{\mP}
\psi^{\dag}_{{\mP}{\cV}}\gamma^5\psi_{{\mP}{\cV}}\right).
\nonumber
\end{gather}
On the other hand, as was stated before, $J=1$. Therefore, since the
quantities $\alpha_{\cV}$ are arbitrary at each vertex, we have
\begin{gather}
\sum_{\mP} \psi^{\dag}_{{\mP}{\cV}}\gamma^5\psi_{{\mP}{\cV}}=0.
\label{d60}
\end{gather}

For the following analysis it is necessary to decompose the sum
(\ref{d60}) into infrared or long-wavelength and the rest
ultraviolet parts. Firstly let's consider the infrared part. One
must introduce the following scales: the gauge field wavelength
order $\sim\lambda$; the scale of ultraviolet cutoff of the
long-wavelength sector $\Lambda$; the lattice scale
$l_P\sim\left|e^a_{{\cW}ij}\right|$. The scales satisfy inequalities
\begin{gather}
\lambda^{-1}\ll\Lambda\ll l_P^{-1}. 
\label{d70}
\end{gather}

Let us divide the total index set $\{{\mP}\}$ into three subsets.
For the long-wavelength $\psi_{{\mP}}(x)$:
\begin{gather}
{\mP}\in{\cal S}_{\mbox{infra}}\quad\Longleftrightarrow\quad |\epsilon_{\mP}|<\Lambda_1, 
\quad \lambda^{-1}\ll\Lambda_1\ll l_P^{-1},
\nonumber \\
{\mP}\in{\cal S}^{\circledcirc}_{\mbox{infra}}\quad\Longleftrightarrow\quad \Lambda_1<|\epsilon_{\mP}|<\Lambda_2\ll l_P^{-1}. 
\nonumber
\end{gather}
The rest of indexes is designated as ${\cI}$, so that
\begin{gather}
{\cal S}_{\mbox{infra}}+{\cal S}^{\circledcirc}_{\mbox{infra}}+{\cI}=\{{\mP}\}
\nonumber
\end{gather}

In  consequence of Eq. (\ref{cor373}) it is evident that for all ${\mP}$ with $\epsilon_{\mP}\neq0$ (see Eq. (\ref{cor363}))
\begin{gather}
\frac15\sum_{\cV}v_{\cV}
\psi^{\dag}_{{\mP}{\cV}}\gamma^5\psi_{{\mP}{\cV}}=0.
\label{d73}
\end{gather}
Due to Eq. (\ref{d73}) and the identity $\gamma^5\equiv\left[(1+\gamma^5)/2-(1-\gamma^5)/2\right]$ we obtain
the relation
\begin{gather}
\frac15\sum_{\cV}v_{\cV}\sum_{\mP\in{\cal S}}
\psi^{\dag}_{{\mP}{\cV}}\gamma^5\psi_{{\mP}{\cV}}=n_+^{\cal S}-n_-^{\cal S},
\label{d74}
\end{gather}
where ${\cal S}$ is a subset of the index set $\{{\mP}\}$ and $n_+^{\cal S}$ $\left(n_-^{\cal S}\right)$
is the number of right (left) zero modes on the index subset ${\cal S}$. In any case the value of the left-hand side 
of Eq. (\ref{d74}) is a whole number $0,\,\pm1,\ldots$.

The value of the long-wavelength part of the sum (\ref{d60}) is well
known \footnote{Note that the expression in the right hand side of Eq. (\ref{d90}) and the integral in the right hand side of Eq. (\ref{d100}) are generalized easily into
irregular lattice (cimplicial complex) in such a way, that the lattice values transform into the corresponding original continual values in the long-wavelength limit.}:
\begin{gather}
\sum_{{\mP}\in{\cal S}_{\mbox{infra}}}
\psi^{\dag}_{{\mP}}(x)\gamma^5\psi_{{\mP}}(x)=
-\frac{e^2}{32\pi^2}\varepsilon^{abcd}\tr\left\{F_{ab}(x)F_{cd}(x)\right\}+
\nonumber \\
+\mbox{O}\left(\frac{1}{(\lambda\Lambda_1)^2}\right){\cal F}_1\{A\}+
\mbox{O}\left(\frac{l_P}{\lambda}\right){\cal F}_2\{A\},
\nonumber \\
F_{ab}=\partial_aA_b-\partial_bA_a+ie\left[A_a,\,A_b\right].
\label{d80}
\end{gather}
Here ${\cal F}_1\{A\}$ and ${\cal F}_2\{A\}$ are some local gauge invariant functionals of the gauge field.
The rigorous lattice expression for the left hand side of Eq. (\ref{d80}) looks like
\begin{gather}
\sum_{{\mP}\in{\cal S}_{\mbox{infra}}}
\psi^{\dag}_{{\mP}{\cV}}\gamma^5\psi_{{\mP}{\cV}}=\tr\gamma^5K_{{\cV},{\cV}}(\Lambda_1),
\nonumber \\
K_{{\cV}_1,{\cV}_2}(\Lambda_1)\equiv
\sum_{\mP}\exp\left[-\frac{(\epsilon_{\mP})^2}{\Lambda_1^2}\right]
\psi_{{\mP}{\cV}_1}\psi^{\dag}_{{\mP}{\cV}_2}=
\nonumber \\  
=\exp\left[-\frac{(i\not\!\!{\cD})^2}{\Lambda_1^2}\right]_{{\cV}_1,{\cV}_2}.
\label{d83}
\end{gather}
The expansion of the lattice operator $K_{{\cV}_1,{\cV}_2}(\Lambda_1)$ in power series in
$(\lambda\Lambda_1)^{-2}\ll1$ and $\left(l_P/\lambda\right)\ll1$ leads to the expression in the right hand
side of Eq. (\ref{d80}). It is important that this expansion is correct since the 
operator $K_{{\cV}_1,{\cV}_2}(\Lambda_1)$
is well defined. 

The space integral of the right-hand side of Eq. (\ref{d80}) is equal to
\begin{gather}
q+\mbox{O}\left(\frac{1}{(\lambda\Lambda)^2}\right)c_1+\mbox{O}\left(\frac{l_P}{\lambda}\right)c_2, 
\nonumber \\
\frac{c_1}{(\lambda\Lambda)^2}\rightarrow0, \quad \frac{l_Pc_2}{\lambda}\rightarrow0.
\label{d85}
\end{gather}
Here $q=0,\pm1,\ldots$ is the topological charge of the gauge field
instanton and the numbers $c_1$ and $c_2$ tend to some finite values in the limit  
$\left(1/\lambda\Lambda\right)\rightarrow0$ and $\left(l_P/\lambda\right)\rightarrow0$. 
Since the value of the left-hand side of Eq. (\ref{d80}) is a whole number (see Eq. (\ref{d74}))
and the latter two summands in (\ref{d85}) are negligible in comparison with $1$, so one must conclude
that $c_1=c_2=0$. Finally we have:
\begin{gather}
\frac15\sum_{\cV}v_{\cV}\sum_{{\mP}\in{\cal S}_{\mbox{infra}}}
\psi^{\dag}_{{\mP}{\cV}}\gamma^5\psi_{{\mP}{\cV}}=q.
\label{d86}
\end{gather}
This equation is rigorous for $\left(1/\lambda\Lambda\right)\lll1$, $\left(l_P/\lambda\right)\lll1$.
Moreover, 
it follows from the Eq. (\ref{d80}), that
\begin{gather}
\sum_{{\mP}\in{\cal S}_{\mbox{infra}}}
\psi^{\dag}_{{\mP}}(x)\gamma^5\psi_{{\mP}}(x)=
\nonumber \\
=-\frac{e^2}{32\pi^2}\varepsilon^{abcd}\tr\left\{F_{ab}(x)F_{cd}(x)\right\}.
\label{d87}
\end{gather}
in the limit $\left(1/\lambda\Lambda\right)\rightarrow0$ and $\left(l_P/\lambda\right)\rightarrow0$. It is well known that the right-hand side of Eq. (\ref{d87}) is a one-half of the
axial vector anomaly. Here the expression for the anomaly is
extracted from the fermion measure (\ref{d30}). This method was
suggested by Vergeles \cite{9} and Fujikawa \cite{10}.

Note that the value of the sum in (\ref{d87}) does not depend on the
cutoff parameter $\Lambda$ if it is enclosed in a range of values
(\ref{d70}). This fact in turn means that
\begin{gather}
\sum_{{\mP}\in{\cal S}^{{\circledcirc}}_{\mbox{infra}}}
\psi^{\dag}_{{\mP}}(x)\gamma^5\psi_{{\mP}}(x)=0. 
\label{d90}
\end{gather}
It is clear from here that the decomposition of the sum in
(\ref{d60}) into long-wavelength and ultraviolet parts is well
defined.

The comparison of Eqs. (\ref{d74}), (\ref{d60}), (\ref{d86}) and (\ref{d90}) leads to the following equality:
\begin{gather}
\frac15\sum_{\cV}v_{\cV}\sum_{{\mP}\in{\cI}}
\psi^{\dag}_{{\mP}{\cV}}\gamma^5\psi_{{\mP}{\cV}}=n^{\cI}_+-n^{\cI}_-=-q.
\label{d110}
\end{gather}
Here $n^{\cI}_+$ ($n^{\cI}_-$) is the number of the right (left)
{\it irregular} zero modes of Eq. (\ref{cor363}). The difference
between the usual and irregular modes is as follows: For the usual
modes and adjacent vertices $a_{{\cW}i}$ and $a_{{\cW}j}$ we have
\begin{gather}
\left|\psi_{({\mP}){\cW}i}-\psi_{({\mP}){\cW}j}\right|\sim
l_P\,\epsilon_{\mP}\left|\psi_{({\mP}){\cW}j}\right|\rightarrow0.
\label{d140}
\end{gather}
By definition, the irregular modes can not satisfy the estimation
(\ref{d140}), but they satisfy the estimation
\begin{gather}
\left|\psi^{\cI}_{({\mP}){\cW}i}-\psi^{\cI}_{({\mP}){\cW}j}\right|
\sim\left|\psi^{\cI}_{({\mP}){\cW}i}\right|
\label{d150}
\end{gather}
at least at a part of vertices. Thus, the usual and irregular modes
are well separated not only by the energy $\epsilon_{\mP}$ but also
by the "momentum".

It is important that the relations (\ref{d110}) are rigorous.

\section{Wilson fermion doubling phenomenon}

Let $q\in{\boldsymbol{\Z}}$ and $\not\!\!\!{\cD}^{(q)}_{{\cV}_1,{\cV}_2}$ be the Dirac operator defined on an instanton
with the topological charge $(q)$. Denote by
$\psi^{\cI}_{(0\,\xi){\cV}}$ the irregular zero
mode of Eq. (\ref{cor363}):
\begin{gather}
\sum_{{\cV}_2}\left[-\frac{i}{v_{{\cV}_1}}\not\!\!{\cD}^{(q)}_{{\cV}_1,{\cV}_2}\right] \psi^{\cI}_{(0\,\xi){\cV}_2}=0.
\label{d160}
\end{gather}
The index $\xi$ enumerates the zero modes.

Now let's denote by
$\left[-\left(i/v_{{\cV}_1}\right)\not\!\!{\cD}^{(\mbox{free})}_{{\cV}_1,{\cV}_2}\right]$
the free lattice Dirac operator. Free Dirac operator is obtained from the general one by the
gauge field elimination: $U_{{\cW}mi}=\exp\left(ieA_{{\cW}mi}\right)\rightarrow 1$

It is easy to obtain the following estimation:
\begin{gather}
\sum_{{\cV}_2}\left[-\frac{i}{v_{{\cV}_1}}\not\!\!{\cD}^{(\mbox{free})}_{{\cV}_1,{\cV}_2}\right]
\psi^{\cI}_{(0\,\xi){\cV}_2}=\mbox{O}\left(\frac{e}{\rho}
\left|\psi^{\cI}_{(0\,\xi){\cV}_1}\right|\right).
\label{d170}
\end{gather}
Here $\rho$ is the scale of the instanton field
$A^{(\mbox{inst})}_{{\cW}mi}$. The proof of (\ref{d170}) is based on
the estimations
\begin{gather}
A^{(\mbox{inst})}_{{\cW}mi}\sim\left(l_P/\rho\right)\ll1,
\nonumber \\
1\approx\exp\left(ieA^{(\mbox{inst})}_{{\cW}mi}\right)-
ieA^{(\mbox{inst})}_{{\cW}mi}=
U_{{\cW}mi}+\mbox{O}\left(\frac{e\,l_P}{\rho}\right),
\nonumber
\end{gather}
and the fact that the lattice Dirac operator is linear in
$U_{{\cW}mi}$. Therefore
\begin{gather}
\not\!\!{\cD}^{(\mbox{free})}_{{\cV}_1,{\cV}_2}=\not\!\!{\cD}^{(q)}_{{\cV}_1,{\cV}_2}+
\mbox{O}\left(\frac{e\,l_P^4}{\rho}\right).
\nonumber
\end{gather}
Since $v_{{\cV}_1}\sim l_P^4$, the estimation (\ref{d170}) follows
from Eq. (\ref{d160}).

Let's expand the field configuration
$ \psi^{\cI}_{(0\,\xi){\cV}}$ in a series of the free Dirac
operator eigenfunctions
\begin{gather}
\psi^{\cI}_{(0\,\xi){\cV}}=\sum_{\mP}c_{\mP}\psi^{(\mbox{free})}_{({\mP}){\cV}},
\nonumber \\
\sum_{{\cV}_2}\left[-\frac{i}{v_{{\cV}_1}}\not\!\!{\cD}^{(\mbox{free})}_{{\cV}_1,{\cV}_2}\right]
\psi^{(\mbox{free})}_{({\mP}){\cV}_2}=\epsilon_{\mP}\psi^{(\mbox{free})}_{({\mP}){\cV}_1}.
\label{d190}
\end{gather}
Here $c_{\mP}$ are some complex numbers.

We are interested in the irregular modes contribution to the
expansion (\ref{d190}):
\begin{gather}
\psi^{\cI}_{(0\,\xi){\cV}}=\sum_{\mP'}
c_{\mP'}\psi^{(\mbox{free})\cI}_{({\mP'}){\cV}}+\ldots\,,
\label{d200}
\end{gather}
where the indices $\mP'$ enumerate the irregular modes. It is
evident that at least some numbers $c_{\mP'}$ in (\ref{d200}) are
nonzero:
\begin{gather}
c_{\mP'}\neq 0.
\label{d210}
\end{gather}
Indeed, the irregular field configuration cannot be expanded in a
series of the regular smooth modes only.

The estimation (\ref{d170}) and expansion (\ref{d200}) allow to do
the final conclusion: the Wilson fermion doubling phenomenon on
irregular 4-Dimensional lattices does exist. Otherwise, the energy
gap of the order of $\epsilon^{\cI}_{\mP}\sim1/l_P$ would be
expected to take place in the sector of all irregular modes of the
free Dirac operator. As was said, in any case the expansion
(\ref{d200}) contains the irregular modes of the operator. Thus, the
additional contributions of the order of $\left(c_{\mP'}/l_P\right)$
would be in the right-hand side of the estimation (\ref{d170}), the
numbers $c_{\mP'}\neq0$. But the right hand side of the estimation
(\ref{d170}) does not depend on the lattice parameter $l_P$. Thus
there are the soft or low energy irregular Dirac modes, the index
$\mP'$ in the expansion (\ref{d200}) enumerates only the soft modes.
The soft irregular eigenfunctions of the free Dirac operator are called
here as doubled fermion modes.

It is necessary to notice, that the suggested approach is valid also
for the regular lattices or partially regular lattices such as
periodic in one dimension and irregular in the rest dimensions.

To prove the existence of Wilson fermion doubling phenomenon on
irregular 3-Dimensional lattices let us consider the Dirac action on
the Cartesian product of a 3-Dimensional simplicial complex $\mK$
and the set of integers ${\boldsymbol{\R}}$. As before, I assume
that the 3-Dimensional simplicial complex is embedded into
3-Dimensional Euclidean space, the vertexes of the complex are
denoted as $a_{\cV}$, the index
${\cV}=1,2,\dots,\,{\mN}\rightarrow\infty$ enumerates the vertices
and the index ${\cW}$ enumerates 3-simplices. Again it is necessary
to use the local enumeration of the vertices $a_{\cV}$ attached to a
given 3-simplex: the all four vertices of a 3-simplex with index
${\cW}$ are enumerated as $a_{{\cW}i}$, $i,\,j,\ldots=1,2,3,4$.
Later the notations with extra index ${\cW}$ indicate that the
corresponding quantities belong to the 3-simplex with index ${\cW}$.
The Levi-Civita symbol with in pairs different indexes
$\varepsilon_{{\cW}lijk}=\pm 1$ depending on whether the order of
vertices $a_{{\cW}l}a_{{\cW}i}a_{{\cW}j}a_{{\cW}k}$ defines the
positive or negative orientation of this 3-simplex. For each
oriented 1-simplex $a_{{\cW}i}a_{{\cW}j}$ of the simplicial complex
an elementary vector
\begin{gather}
e^{\alpha}_{{\cW}ij}\equiv-e^{\alpha}_{{\cW}ji},  \quad
\alpha,\,\beta,\,\gamma=1,\,2,\,3
\nonumber
\end{gather}
is assigned. The vector $e^{\alpha}_{{\cW}ij}$ connects the vertex
$a_{{\cW}i}$ with the vertex $a_{{\cW}j}$ in 3D Euclidean space. The
rest of notations are evident and they are similar to that in the
beginning of Section 2, but they are
supplied here by the additional index
$n=0,\,\pm1,\ldots\in{\boldsymbol{\R}}$ since the dynamic variables are defined
now on the discrete set $\mK\times{\boldsymbol{\R}}$.

The Euclidean Hermitean action of the Dirac field associated with
the set $\mK\times{\boldsymbol{\R}}$ has the form
\begin{widetext}
\begin{gather}
\mA_{\psi}=-\frac{1}{2!\,4!)}
\sum_n\sum_{\cW}\sum_{i,j,k,l}\varepsilon_{{\cW}lijk}\,\varepsilon^{\alpha\beta\gamma}
\left(i\,\psi_{{\cW}l,\,n}^{\dag}\gamma^{\alpha}\psi_{{\cW}i,\,n}\right)
e^{\beta}_{{\cW}lj,\,n}e^{\gamma}_{{\cW}lk,\,n}-
\frac{1}{2}\sum_n\sum_{\cV}v_{\cV}\left(i\,\psi_{{\cV},\,n}^{\dag}\gamma^4
(\psi_{{\cV},\,n+1}-\psi_{{\cV},\,n-1})\right)=
\nonumber \\
=\sum_n\sum_{{\cV}_1{\cV}_2}\psi^{\dag}_{{\cV}_1n}\left[
-i\gamma^{\alpha}{\cD}^{\alpha}_{{\cV}_1,{\cV}_2}
\right]\psi_{{\cV}_2n}+\sum_{\cV}v_{\cV}
\sum_{n,n'}\psi_{{\cV},\,n}^{\dag}\left[-i\,\gamma^4D_{n,n'}\right]
\psi_{{\cV},\,n'}.
\label{d220}
\end{gather}
\end{widetext}
Here $v_{\cV}$ is the total sum of oriented volumes of the adjacent
3-simplices with common vertex $a_{\cV}$. The eigenfunction problem
(\ref{cor363}) for irregular modes now looks like
\begin{widetext}
\begin{gather}
\sum_{{\cV}_2}\left[-\frac{i}{v_{{\cV}_1}}{\gamma^{\alpha}
\cD}^{\alpha}_{{\cV}_1,{\cV}_2}\right] \psi^{\cI}_{({\mP}){\cV}_2n}+
\sum_{n'}\left[-i\,\gamma^4D_{n,n'}\right]\psi^{\cI}_{({\mP}){\cV}_1n'}
=\epsilon_{\mP}\psi^{\cI}_{({\mP}){\cV}_1n},
\label{d230}
\end{gather}
\end{widetext}
or briefly
\begin{gather}
\left\{\gamma^{\alpha}{(-i/v_{{\cV}})\cD}^{\alpha}+\gamma^4(-iD)\right\}
\psi^{\cI}_{({\mP})}=\epsilon_{\mP}\psi^{\cI}_{({\mP})}.
\label{d240}
\end{gather}
Both operators ${(-i/v_{{\cV}_1})\cD}^{\alpha}_{{\cV}_1,{\cV}_2}$
and $-iD_{n,n'}$ are Hermitean and they commute mutually. Therefore, the repeated
application of the operator
$\left\{\gamma^{\alpha}{(-i/v_{{\cV}_1})\cD}^{\alpha}+\gamma^4(-iD)\right\}$
to (\ref{d240}) leads to the equation
\begin{gather}
\left\{\left[(i/v_{{\cV}}){\cD}^{\alpha}\right]^2+\left[iD\right]^2\right\}
\psi^{\cI}_{({\mP})}=\epsilon^2_{\mP}\psi^{\cI}_{({\mP})}.
\label{d250}
\end{gather}
due to the fact that
$\gamma^{\alpha}\gamma^4+\gamma^4\gamma^{\alpha}=0$, It has been shown that the soft irregular modes of Eqs. (\ref{d240})
and (\ref{d250}) do exist, i.e. there exist the eigenvalues of the
Eq. (\ref{d240}) in the subspace of irregular eigenfunctions of the
order of $|\epsilon_{\mP}|\ll l_P^{-1}$. Therefore the spectrum of
the operator $\left[(i/v_{{\cV}}){\cD}^{\alpha}\right]$ in the
subspace of irregular eigenfunctions contains the eigenvalues of the
order of $|\epsilon_{\mP}|\ll l_P^{-1}$. This conclusion follows
from Eq. (\ref{d250}).

Thus, the doubled fermion modes exist also on 3-Dimensional
irregular lattices.

The classification of the doubled fermion modes should be a subject
of future scientific research.

\section{The propagation of the irregular quanta}

At first let us fix the necessary properties of the usual Dirac
propagators
\begin{gather}
iS_c(x-y)\equiv\langle0|T\psi(x)\opsi(y)|0\rangle
\label{pr10}
\end{gather}
in $(3+1)$ continual space-time with Minkowski signature:

1) the translational and Lorentz invariance;

2) for massless theory
\begin{gather}
\gamma^5iS_c(x-y)+iS_c(x-y)\gamma^5=0; \label{pr20}
\end{gather}

3) for $x^0>z^0>y^0$
\begin{gather}
\int\d^{(3)}z\left[iS_c(x-z)\right]\gamma^0\left[iS_c(z-y)\right]
=iS_c(x-y);
\label{pr30}
\end{gather}

4) the propagating particles are "good" quasiparticles, i.e. they
live indefinitely and have well defined four-momentum and their
energy is positive.

The property 3) is the quantum-mechanical superposition principle
and  at the same time the property implies that the propagating
particle can not be absorbed or created by vacuum, i.e. the particle
is distinguishable against the background of the vacuum.

It is easy to see that all four of the properties define uniquely
the particle propagator. Indeed, the most general expression for the
propagator in the case $x^0>y^0$ is
\begin{gather}
iS_c(x-y)=\int\left(\frac{\d^{(3)}k}{(2\pi)^32|{\bf
k}|}\right)\times
\nonumber \\
\times\left(\gamma^0|{\bf k}|-\gamma^{\alpha}k^{\alpha}\right)
e^{i{\bf k}({\bf x}-{\bf y})-i|{\bf k}|(x^0-y^0)}f(k^a).
\label{pr40}
\end{gather}
Here the measure, the expression in the parentheses and the exponent
are Lorentz-invariant. The property 2) is also fulfilled. Since the
propagator (\ref{pr40}) describes the propagation of the real "good"
quasiparticles, so the all its dependence on the space-time
coordinates $(x-y)$ is given by the exponent. The function $f(k^a)$
in (\ref{pr40}) also must be Lorentz-invariant. This means that it
can depend only on $k^ak_a=0$ and thus it is constant: $f=C$. The
property 3) gives $C^2=C$. Therefore $f(k^a)=1$.

If we insist on the properties 1)-2) only and reject the properties
3)-4), then the propagator describes the propagation of some
irregular quanta and it can acquire another forms. For example
\begin{gather}
iS^{\cI}_c(x-y)\sim l_P^2i\gamma^a\left(\partial/\partial
x^a\right)\delta^{(4)}(x-y).
\label{pr50}
\end{gather}

It is shown below that the propagators of the irregular quanta are
similar to the expression (\ref{pr50}).  In order to do this, the
structure of the fermion vacuum must be described in  general.

Now I return to the Euclidean metric. For simplicity, the gauge
group is assumed to be trivial, so that the index $\varkappa$ will
be omitted. Note that from the integration rules
\begin{gather}
\int\d\psi_{{\cV} s}=0, \quad \int\d\psi_{{\cV} s}\cdot\psi_{{\cV}
s'}=\delta_{s\,s'},  
\nonumber \\
\int\d\psi^{\dag}_{{\cV} s}=0, \quad
\int\d\psi^{\dag}_{{\cV} s}\cdot\psi^{\dag}_{{\cV}
s'}=\delta_{s\,s'} \label{pr60}
\end{gather}
it follows that the nonzero value of the integral (\ref{d10}) is
obtained only if the complete products of the fermion variable
\begin{gather}
\left(\prod_{s=1}^4\psi_{{\cV} s}\psi^{\dag}_{{\cV} s}\right)
\label{pr70}
\end{gather}
are present at each vertex $a_{\cV}$. These products can arise only
due to the exponent expansion under the integral (\ref{d10}). As a
consequence of the expansion the expression
$\left\{\psi^{\dag}_{{\cV}_1s_1}
\left[-i\not\!\!{\cD}_{{\cV}_1,{\cV}_2}\right]_{s_1s_2}\psi_{{\cV}_2s_2}\right\}$
related to the 1-simplex $a_{{\cV}_1}a_{{\cV}_2}$ can appear (see
the Dirac action (\ref{dqg110})) \footnote{By definition of the
matrix $\left[-i{\cD}_{{\cV}_1,{\cV}_2}\right]_{s_1s_2}$ the indices
${\cV}_1$ and ${\cV}_2$ enumerate the nearest vertices $a_{{\cV}_1}$
and $a_{{\cV}_2}$, i.e. the vertices belonging to the same 1-simplex
$a_{{\cV}_1}a_{{\cV}_2}$.}. Let's assign to the
corresponding 1-simplex $a_{{\cV}_1}a_{{\cV}_2}$ an arrow in this case. The arrow
is vectored from vertex $a_{{\cV}_2}$ to vertex $a_{{\cV}_1}$ which
can be designated as $\overrightarrow{a_{{\cV}_2}a_{{\cV}_1}}$ or
$\overleftarrow{a_{{\cV}_1}a_{{\cV}_2}}$. Four arrows come into each vertex and four arrows
come out from each vertex as a result of integration
in (\ref{d10}). This geometrical picture is realized
analytically by assigning to each 1-simplex
$\overleftarrow{a_{{\cV}}a_{{\cV}_1}}$ the matrix
$\left[-i\not\!\!{\cD}_{{\cV},{\cV}_1}\right]_{s\,s_1}$ and to each
1-simplex $\overrightarrow{a_{{\cV}}a_{{\cV}_1}}$ the matrix
$\left[-i\not\!\!{\cD}_{{\cV}_1,{\cV}}\right]_{s_1s}$. Thus there is the factor
\begin{widetext}
\begin{gather}
\left\{\sum_{s_1,s_2,s_3,s_4=1}^4\varepsilon_{s_1s_2s_3s_4}
\left[-i\not\!\!{\cD}_{{\cV},{\cV}_1}\right]_{s_1s'_1}
\left[-i\not\!\!{\cD}_{{\cV},{\cV}_2}\right]_{s_2s'_2}
\left[-i\not\!\!{\cD}_{{\cV},{\cV}_3}\right]_{s_3s'_3}
\left[-i\not\!\!{\cD}_{{\cV},{\cV}_4}\right]_{s_4s'_4}\right\}\times
\nonumber \\
\times\left\{\sum_{s_5,s_6,s_7,s_8=1}^4\varepsilon_{s_5s_6s_7s_8}
\left[-i\not\!\!{\cD}_{{\cV}_5,{\cV}}\right]_{s'_5s_5}
\left[-i\not\!\!{\cD}_{{\cV}_6,{\cV}}\right]_{s'_6s_6}
\left[-i\not\!\!{\cD}_{{\cV}_7,{\cV}}\right]_{s'_7s_7}
\left[-i\not\!\!{\cD}_{{\cV}_8,{\cV}}\right]_{s'_8s_8}\right\}.
\label{pr80}
\end{gather}
\end{widetext}
in every vertex $a_{\cV}$ 

We are interested in the two-point correlator
\begin{gather}
\langle\psi_{{\cV}_1s_1}\psi^{\dag}_{{\cV}_2s_2}\rangle\equiv
\frac{\int\left({\cD}\psi^{\dag}{\cD}\psi\right)
\psi_{{\cV}_1s_1}\psi^{\dag}_{{\cV}_2s_2}\exp\mA_{\psi}}
{\int\left({\cD}\psi^{\dag}{\cD}\psi\right) \exp\mA_{\psi}}.
\label{pr90}
\end{gather}
Since there is the external factor
$\psi^{\dag}_{{\cV}_2s_2}$ in the vertex $a_{{\cV}_2}$, the number of the arrows related with
the factors
\begin{gather}
\left[-i\not\!\!{\cD}_{{\cV}_2,{\cV}'}\right]_{s_2s'}
\label{pr100}
\end{gather}
and coming into the vertex $a_{{\cV}_2}$ is reduced up to tree.
Mathematically this fact is realized by the assigning the inverse
matrix
\begin{gather}
\left[-i\not\!\!{\cD}_{{\cV}_2,{\cV}'}\right]^{-1}_{s's_2}, 
\nonumber \\
\sum_{s'}\left[-i\not\!\!{\cD}_{{\cV}_2,{\cV}'}\right]^{-1}_{s_1s'}
\left[-i\not\!\!{\cD}_{{\cV}_2,{\cV}'}\right]_{s's_2}=\delta_{s_1,s_2}
\label{pr110}
\end{gather}
to the corresponding 1-simplex $a_{{\cV}_2}a_{{\cV}'}$ (see Fig. 1).
Therefore the number of factors $\psi_{{\cV}'s'}$ presented at the
vertex $a_{{\cV}'}$ is reduced up to tree also. To compensate this
reduction one must introduce the additional factor (see Fig. 1)
\begin{gather}
\left[-i\not\!\!{\cD}_{{\cV}'',{\cV}'}\right]_{s''s'}.
\label{pr120}
\end{gather}

\begin{figure}[t]
\psfrag{fp1}{\rotatebox{0}{\kern-10pt\lower0pt\hbox{{$\psi_
{{\cal{V}}_1s_1}$}}}}
\psfrag{fp2}{\rotatebox{0}{\kern0pt\lower0pt\hbox{{${\cal{V}}_1$}}}}
\psfrag{fp3}{\rotatebox{22}{\kern-16pt\lower0pt\hbox{{additional}}}}
\psfrag{fp4}{\rotatebox{22}{\kern-12pt\lower2pt\hbox{{reduced}}}}
\psfrag{fp5}{\rotatebox{0}{\kern0pt\lower-3pt\hbox{{${\cal{V}}'''''$}}}}
\psfrag{fp6}{\rotatebox{-22}{\kern-16pt\lower0pt\hbox{{additional}}}}
\psfrag{fp7}{\rotatebox{0}{\kern0pt\lower0pt\hbox{{${\cal{V}}''''$}}}}
\psfrag{fp8}{\rotatebox{0}{\kern5pt\lower0pt\hbox{{${\cal{V}}'''$}}}}
\psfrag{fp9}{\rotatebox{-90}{\kern-27pt\lower-1pt\hbox{{reduced}}}}
\psfrag{fp01}{\rotatebox{-90}{\kern-33pt\lower-1pt\hbox{{additional}}}}
\psfrag{fp02}{\rotatebox{0}{\kern0pt\lower0pt\hbox{{${\cal{V}}''$}}}}
\psfrag{fp03}{\rotatebox{22}{\kern-12pt\lower4pt\hbox{{additional}}}}
\psfrag{fp04}{\rotatebox{0}{\kern2pt\lower0pt\hbox{{${\cal{V}}'$}}}}
\psfrag{fp05}{\rotatebox{-22}{\kern-15pt\lower-1pt\hbox{{additional}}}}
\psfrag{fp06}{\rotatebox{-22}{\kern-12pt\lower-2pt\hbox{{reduced}}}}
\psfrag{fp07}{\rotatebox{0}{\kern-4pt\lower0pt\hbox{{$\psi^+_
{{\cal{V}}_2s_2}$}}}}
\psfrag{fp08}{\rotatebox{0}{\kern0pt\lower0pt\hbox{{${\cal{V}}_2$}}}}
\psfrag{fp09}{\rotatebox{0}{\kern0pt\lower0pt\hbox{{vertex}}}}
\psfrag{pf1}{\rotatebox{0}{\kern0pt\lower2pt\hbox{{$\left[-i\not\!\!{\cal{D}}_
{{\cal{V}}_2,{\cal{V}}_1}\right]_{s_2s_1}$}}}}
\psfrag{pf2}{\rotatebox{0}{\kern0pt\lower2pt\hbox{{$\left[-i\not\!\!{\cal{D}}_
{{\cal{V}}_2,{\cal{V}}_1}\right]^{-1}_{s_1s_2}$}}}}
    \center{\includegraphics[width=0.40\textwidth]{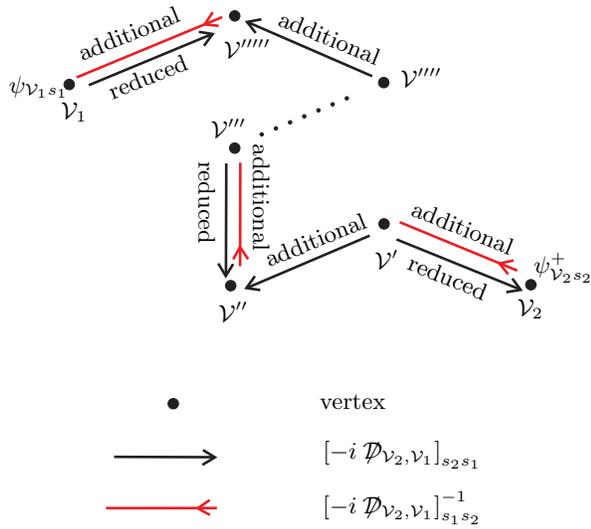}}
    \caption{The graphical representation of the curly brackets in the right-hand side of Eq. (\ref{pr130})}
    \label{fig:1}
\end{figure}

Now the condition at the vertex $a_{{\cV}''}$ is the same as at the beginning of the process at the
vertex $a_{{\cV}2}$: the additional
factor (\ref{pr120}) gives an additional arrow coming into the vertex
$a_{{\cV}''}$. To eliminate one of them, say
$\overleftarrow{a_{{\cV}''}a_{{\cV}'''}}$, one should introduce the
factor $\left[-i\not\!\!{\cD}_{{\cV}'',{\cV}'''}\right]^{-1}_{s'''s''}$, and
so on. It is evident that the last link in the chain is
$\left[-i\not\!\!{\cD}_{{\cV}'''',{\cV}_1}\right]^{-1}_{s_1s''''}$.

It follows from the above-said that the correlator (\ref{pr90}) can
be represented in the form
\begin{widetext}
\begin{gather}
\langle\psi_{{\cV}_1s_1}\psi^{\dag}_{{\cV}_2s_2}\rangle=\sum_{\mbox{all
paths}}\bigg\{\left[-i\not\!\!{\cD}_{{\cV}''''',{\cV}_1}\right]^{-1}
\left[-i\not\!\!{\cD}_{{\cV}''''',{\cV}''''}\right]\ldots
\left[-i\not\!\!{\cD}_{{\cV}'',{\cV}'''}\right]^{-1}
\left[-i\not\!\!{\cD}_{{\cV}'',{\cV}'}\right]
\left[-i\not\!\!{\cD}_{{\cV}_2,{\cV}'}\right]^{-1}\bigg\}_{s_1s_2}.
\label{pr130}
\end{gather}
\end{widetext}
Obviously, the number of
the operators $\left[-i{\cD}\right]^{-1}$ is greater than the number
of the operators $\left[-i\not\!\!{\cD}\right]$ by the unity in the right-hand side of Eq. (\ref{pr130}). Therefore the
total power of the operators $\left[-i\not\!\!{\cD}\right]$ and
$\left[-i\not\!\!{\cD}\right]^{-1}$ in the right-hand side of Eq.
(\ref{pr130}) is odd. Since both these operators are linear in the
Dirac matrices $\gamma^a$, so the expression in the right-hand side
of Eq. (\ref{pr130}) satisfies the property 2). But the property 3)
can not be fulfilled on the microscopic level -  if only because of
the correlator (\ref{pr90}) is odd in the total power of the Dirac
matrices while the bilinear form of the correlator is even in this
sense. Note that a part of information is lost in passing from the microscopic description to the
long wavelength limit, and thus the
property 3) becomes true. Indeed, the
information about the lattice is lost completely in the long wavelength limit and the lattice
action (\ref{dqg110}) transforms to the usual continuum Dirac action
(\ref{cor385}). Therefore the correlator (\ref{pr90}) transforms to
the expression (\ref{pr40}) with $f(k^a)=1$.

Now let's proceed to the estimation of the irregular quanta
correlator. In this case the information related with the lattice is
determinative. Because of this, Eq. (\ref{pr130}) should be used.
Since the direct correlator estimation with the help of Eq.
(\ref{pr130}) is impossible, I apply a simple and adequate
computational model which describes the problem in terms of
continuum theory. Thus the model forgets the details of the lattice.

It is supposed here that the microscopic geometry of the lattice is
not fixed. This means that the elementary vectors (\ref{dqg100})
connecting the nearest vertices $a_{{\cW}i}$ and $a_{{\cW}j}$ are
quantum variables, so that their quantum fluctuations are described
by the corresponding wave function. This point of view is necessary
in the lattice quantum theory of gravity \cite{7}-\cite{8}. Though
this theory is not satisfactory at present, I hold to the following
point of view: if the space-time is discrete on microscopic level,
then the corresponding lattice is irregular and the geometrical
values describing the lattice are quantum variables. Such lattice is
called as "breathing" one.

It seems that the propagation of an irregular fermion on the
considered "breathing" lattice is similar in a sence to the dynamics
of a Brownian particle: in the process of successive movements of
fermion from one vertex to another the information of a previous
jump is forgotten due to the irregularity and "breathing" of the
lattice. Thus the propagation of irregular fermions can be described
by a slightly modified Markov process which must  model the
correlator (\ref{pr130}) in the 4-Dimensional Euclidean space.

It is seen from Eqs. (\ref{dqg110}) and (\ref{cor354}) that
\begin{gather}
\sum_{a=1}^4e^a_{{\cV}_1,{\cV}_2}{\cD}^a_{{\cV}_1,{\cV}_2}\sim
v_{{\cV}_1,{\cV}_2},
\label{pr140}
\end{gather}
is the sum of oriented volumes of all 4-simplexes with the common
1-simplex $a_{{\cV}_1}a_{{\cV}_2}$. Therefore, the model of the
amplitude $\left[-i\gamma^a{\cD}^a_{{\cV}_1,{\cV}_2}\right]$ in
(\ref{pr130}) will be the following one:
\begin{gather}
\left[-i\not\!\!{\cD}_{{\cV}_1,{\cV}_2}\right]\longrightarrow
\nonumber \\
\left[-i\not\!\!{\cD}(x-y)\right]\equiv
\left[\frac{\rho}{\pi
b}\left(-i\gamma^a\partial_a\right)
\exp\left(-\frac{(x-y)^2}{b^2}\right)\right].
\label{pr150}
\end{gather}
The right-hand side of (\ref{pr150}) is the amplitude of the jump
from the point $x$ into the point $y$. Here the dimensionless
Cartesian coordinates $x^a\rightarrow x^a/l_P$ are used. The
numerical constant $b\sim1$ is a parameter of the model, $\rho$ is
an unknown normalization constant which is of no importance. It is
seen that the direction of the jump vector $(y-x)$ is unconstrained,
but the jump step value is constrained by the Gauss distribution.
The model of the inverse amplitude
$\left[-i\not\!\!{\cD}_{{\cV}_1,{\cV}_2}\right]^{-1}$ is as follows:
\begin{gather}
\left[-i\not\!\!{\cD}_{{\cV}_1,{\cV}_2}\right]^{-1}\longrightarrow
\nonumber \\
\left[-i\not\!\!{\cD}(x-y)\right]^{-1}\equiv\left[\frac{1}{\pi\rho\,
b}\left(-i\gamma^a\partial_a\right)
\exp\left(-\frac{(x-y)^2}{b^2}\right)\right].
\label{pr160}
\end{gather}
Now the analog of the relation (\ref{pr110}) is the equality
\begin{gather}
\int\d^{(4)}y\left[-i\not\!\!{\cD}(x-y)\right]^{-1}\left[-i\not\!\!{\cD}(y-x)\right]=1.
\label{pr170}
\end{gather}
Thereby the model of the correlator representation (\ref{pr130})
looks like $(z_0=y)$
\begin{gather}
\langle\psi(x)\psi^{\dag}(y)\rangle^{\cI}=\sum_{k=0}^{\infty}\prod_{i=1}^{2k+1}\left\{\int\d^{(4)}z_i\right\}
\delta^{(4)}(x-z_{2k+1})
\nonumber \\
\left[-i\not\!\!{\cD}(z_{2k+1}-z_{2k})\right]^{-1}\left[-i\not\!\!{\cD}(z_{2k}-z_{2k-1})\right]
\ldots
\nonumber \\
\ldots\left[-i\not\!\!{\cD}(z_3-z_2)\right]^{-1}
\left[-i\not\!\!{\cD}(z_2-z_1)\right] \left[-i\not\!\!{\cD}(z_1-y)\right]^{-1}.
\nonumber
\end{gather}
Since the operators $[-i\not\!\!{\cD}]$  and  $\left[-i\not\!\!{\cD}\right]^{-1}$ are coupled one
can put $\rho=1$. This expression is rewritten by passing to the new
integration variables $\tilde{z}_i=z_i-z_{i-1},\,\,i=1,\ldots,2k+1$:
\begin{widetext}
\begin{gather}
\langle\psi(x)\psi^{\dag}(0)\rangle^{\cI}=\sum_{k=0}^{\infty}\prod_{i=1}^{2k+1}\int\d^{(4)}z_i
\delta^{(4)}\left(x-\sum_{j=1}^{2k+1}z_j\right)
\left[-i\not\!\!{\cD}(z_{2k+1})\right]^{-1}\left[-i\not\!\!{\cD}(z_{2k})\right]
\ldots\left[-i\not\!\!{\cD}(z_2)\right]
\left[-i\not\!\!{\cD}(z_1)\right]^{-1}.
\nonumber
\end{gather}
\end{widetext}
With the help of Eqs. (\ref{pr150}) and (\ref{pr160}) the right-hand
side of the last relation is rewritten once again:
\begin{widetext}
\begin{gather}
\langle\psi(x)\psi^{\dag}(0)\rangle^{\cI}=\sum_{k=0}^{\infty}\int\ldots\int
\d^{(4)}z_1\ldots\d^{(4)}z_{2k+1}\delta^{(4)}\left(\sum_{i=1}^{2k+1}z_i-x\right)
\prod_{i=1}^{2k+1}\left[\frac{1}{\pi
b}\left(-i\gamma^a\partial_a\right)
\exp\left(-\frac{z_i^2}{b^2}\right)\right]=
\nonumber \\
=\sum_{k=0}^{\infty}\int\frac{\d^{(4)}q}{(2\pi)^4}e^{-iqx}\prod_{i=1}^{2k+1}
\left[\frac{2}{\pi
b^3}\int(i\gamma^az^a_i)\exp\left(-\frac{z_i^2}{b^2}+iqz_i\right)\d^{(4)}z_i\right]=
\nonumber \\
=\int\frac{\d^{(4)}q}{(2\pi)^4}e^{-iqx}\sum_{k=0}^{\infty}\left[2\pi
b\left(\gamma^a\frac{\partial}{\partial
q^a}\right)\exp\left(-\frac{q^2b^2}{4}\right)\right]^{2k+1}=
\left(-i\gamma^a\frac{\partial}{\partial
x^a}\right)
\int\frac{\d^{(4)}q}{(2\pi)^4}\frac{\pi b^3\exp\left(
-\frac{q^2b^2}{4}-iqx\right)}{1-\pi^2b^6q^2\exp\left(-\frac{q^2b^2}{2}\right)}.
\label{pr180}
\end{gather}
\end{widetext}
Integral in the right-hand side of Eq. (\ref{pr180}) is determined
for
\begin{gather}
0<b<\left(\frac{e}{2\pi^2}\right)^{1/4}\approx0,61.
\label{pr190}
\end{gather}
Integration over the angle variables leads to the expression
$(r\equiv|x|)$
\begin{gather}
\langle\psi(x)\psi^{\dag}(0)\rangle^{\cI}=
\left(-i\gamma^a\frac{\partial}{\partial
x^a}\right)
\nonumber \\
\left[\left(\frac{b^3}{4\pi r}\right) \int_0^{\infty}\d
q\cdot q^2\frac{J_1(qr)\exp\left(
-\frac{q^2b^2}{4}\right)}{1-\pi^2b^6q^2\exp\left(-\frac{q^2b^2}{2}\right)}\right].
\label{pr200}
\end{gather}
The characteristic value of the variable $q$ saturating the integral
(\ref{pr200}) is determined by the nearest zero of the denominator
in the integral. So $|q|\sim1$. Since we are interested in the
correlator behavior for $r\gg1$, the argument $qr$ of the Bessel
function under the integral (\ref{pr200}) is effectively large:
$qr\gg1$. Therefore one can use the asymptotic behavior of the
Bessel function:
\begin{gather}
J_1(qr)\rightarrow\frac{1}{\sqrt{2\pi qr}}\left[e^{iqr-3\pi
i/4}+e^{-iqr+3\pi i/4}\right].
\nonumber
\end{gather}
With the help of the last relation the integral (\ref{pr200}) is
rewritten as follows:
\begin{widetext}
\begin{gather}
\langle\psi(x)\psi^{\dag}(0)\rangle^{\cI}=
\left(-i\gamma^a\frac{\partial}{\partial x^a}\right)
\left[\frac{b^3}{2(2\pi r)^{3/2}} \int_C\d q\cdot
q^{3/2}\frac{\exp\left( -\frac{q^2b^2}{4}+iqr-3\pi
i/4\right)}{1-\pi^2b^6q^2\exp\left(-\frac{q^2b^2}{2}\right)}\right].
\label{pr210}
\end{gather}
\end{widetext}
The integration contour $C$ is pictured on Fig. 2.

\begin{figure}[t]
\psfrag{up01}{\rotatebox{0}{\kern0pt\lower-2pt\hbox{{$q=0$}}}}
\psfrag{up02}{\rotatebox{0}{\kern2pt\lower-1pt\hbox{{$q$}}}}
    \center{\includegraphics[width=0.30\textwidth]{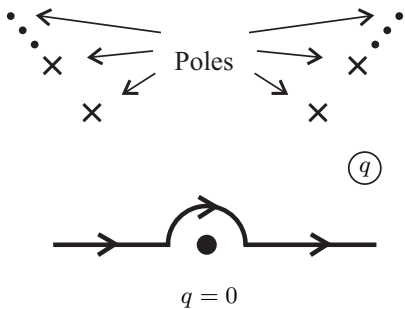}}
    \caption{The integration contour in the integral (\ref{pr210}) and
    the location of the integral poles in the complex plane of $q$-variable}
    \label{fig:1}
  \end{figure}

  \begin{figure}[t]
\psfrag{nuie}{\rotatebox{0}{\kern0pt\lower0pt\hbox{{$\nu
_e^{\cal{I}}$}}}}
\psfrag{nue}{\rotatebox{0}{\kern0pt\lower0pt\hbox{{$\nu_{e}$}}}}
\psfrag{nuimu}{\rotatebox{0}{\kern0pt\lower0pt\hbox{{$\nu
_{\mu}^{\cal{I}}$}}}}
\psfrag{numu}{\rotatebox{0}{\kern0pt\lower0pt\hbox{{$\nu_{\mu}$}}}}
\psfrag{x}{\rotatebox{0}{\kern0pt\lower0pt\hbox{{$x$}}}}
\psfrag{y}{\rotatebox{0}{\kern0pt\lower0pt\hbox{{$y$}}}}
\psfrag{z}{\rotatebox{0}{\kern0pt\lower0pt\hbox{{$z$}}}}
\psfrag{w}{\rotatebox{0}{\kern0pt\lower0pt\hbox{{$w$}}}}
\psfrag{xinf}{\rotatebox{0}{\kern0pt\lower0pt\hbox{{$x_{\infty}$}}}}
\psfrag{yinf}{\rotatebox{0}{\kern0pt\lower0pt\hbox{{$y_{\infty}$}}}}
\psfrag{bp}{\rotatebox{0}{\kern0pt\lower0pt\hbox{{$\bar{\psi}$}}}}
\psfrag{rl}{\rotatebox{0}{\kern0pt\lower0pt\hbox{{$\circlearrowright$}}}}
\psfrag{rp}{\rotatebox{0}{\kern0pt\lower0pt\hbox{{$\circlearrowleft$}}}}
\psfrag{lwf}{\rotatebox{0}{\kern0pt\lower0pt\hbox{{long-wavelength
fermions}}}}
\psfrag{2}{\rotatebox{0}{\kern0pt\lower0pt\hbox{{$2$}}}}
\psfrag{3}{\rotatebox{0}{\kern0pt\lower0pt\hbox{{$3$}}}}
\psfrag{4}{\rotatebox{0}{\kern0pt\lower0pt\hbox{{$4$}}}}
\psfrag{1}{\rotatebox{0}{\kern0pt\lower0pt\hbox{{$1$}}}}
\psfrag{irf}{\rotatebox{0}{\kern0pt\lower0pt\hbox{{irregular
fermions}}}}
\psfrag{gr}{\rotatebox{0}{\kern0pt\lower0pt\hbox{{graviton}}}}
\psfrag{B8}{\rotatebox{0}{\kern0pt\lower0pt\hbox{{$\alpha_{\ve{x}{+}\ve{e}_3}$}}}}
\psfrag{B9}{\rotatebox{0}{\kern0pt\lower0pt\hbox{{$\gamma_{\ve{x}{+}\ve{e}_1}$}}}}
    \center{\includegraphics[width=0.35\textwidth]{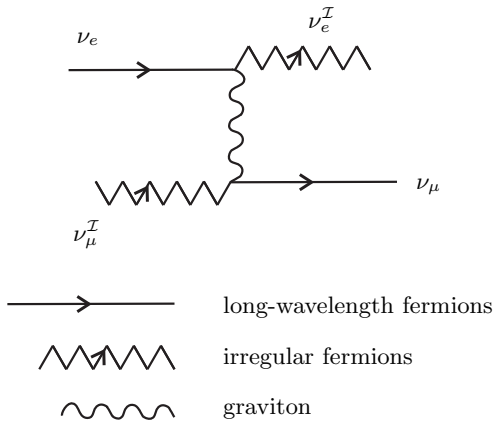}}
    \caption{The process of the electron neutrino transition to the muon one.}
    \label{fig:2}
  \end{figure}

We are interested in the denominator zeros in the upper half plane
of the complex variable $q=q'+iq''$. The zeros are determined by the
following set of equations:
\begin{gather}
(q'^2-q''^2)=2q'q''\ctg(b^2q'q''),
\nonumber \\
2\pi^2b^4\exp[-(b^2q'q'')\ctg(b^2q'q'')]=\frac{\sin(b^2q'q'')}{b^2q'q''}.
\label{pr220}
\end{gather}
Since the solutions of the set of equations (\ref{pr220}) are
symmetrized relative to the imaginary axis, it is enough to solve
the system for $q'>0, \ \ q''>0$. The approximative solution of the
last set of equations looks like
\begin{gather}
b^2q'q''\approx\left(2n+1/2\right)\pi, \quad n=0,1,\ldots,
\nonumber \\
q'_n\sim q''_n\approx\frac{\sqrt{\left(2n+1/2\right)\pi}}{b}.
\label{pr230}
\end{gather}

All zeros of the denominator under the integral (\ref{pr210}) lead
to the simple poles of the expression under the integral sign.
Indeed, the derivative of the denominator respect to the integration
variable is equal to zero only for $q=0,\pm\sqrt{2}/b$. Therefore
\begin{gather}
\mbox{The denominator}=c_n(q-q_n) \ \ \mbox{at} \ \ q\rightarrow
q_n.  \nonumber
\end{gather}
Thus contour $C$ in the integral (\ref{pr210}) can de deformed up,
so that the integral becomes a sum over poles residue. The sum is
saturated by the pair of poles which are nearest to the real axis
 and placed at $q'=\pm\kappa'/b,
\ \ q''=\kappa/b$, where $\kappa', \, \kappa\sim1$ ($n=0$ in
(\ref{pr230})).

Finally we have:
\begin{gather}
\langle\psi(x)\psi^{\dag}(0)\rangle^{\cI}\sim
\left(i\gamma^a\frac{\partial}{\partial
x^a}\right)\left[\frac{1}{r^{3/2}}\exp(-\kappa r/b)\cos\frac{\kappa'
r}{b}\right].
\label{pr240}
\end{gather}
The right-hand side of the relation (\ref{pr50}) simulates the
obtained result (\ref{pr240}) in Minkowski space-time with restored
dimensionality.

We see that the irregular quanta are "bad" quasiparticles.

The fermion lines, such as in Fig. 1, represent creation (at
${\cV}_2$) , propagation and annihilation (at ${\cV}_1$) of a
fermion quantum, and the quantum creation and annihilation events
are induced by the external sources only. If the fermion line is
everywhere continuous and endless in the space then it describes the
propagation of a real particle.

\section{The motivation and speculations}

Instead of a conclusion I would like to present briefly the
motivational factor for the irregular doubled quanta study. 

Let us suppose that the space-time is discrete on microscopic level, the
corresponding lattice is irregular and "breathing" one (see the
previous Section). Suppose also that there are
the nonzero densities of the irregular quanta of the three known neutrinos. 

The nonzero densities $(n^{{\cal I}}\neq 0)$ of the irregular quanta does not contradict
the fundamental notions of astrophysics since the irregular quanta
energy can be arbitrarily small (see the end of Section IV).

The following consequences might have resulted from the suppositions. 

\ \ 

I. {\it{The problem of dark matter in cosmology}}.

Does the nonzero densities of the neutrino irregular quanta form dark matter in cosmology? It seems
that this hypothesis does not contradict to the main properties of dark matter:  
(i) the irregular quanta are "bad" quasiparticles, so such dark matter can be localized; (ii) the irregular quanta interact very slightly with all normal quanta. 

But the nonzero densities of the neutrino irregular quanta give a 
contribution to the energy-momentum tensor and therefore to the gravitational potential in the vicinity of a metagalaxy. 

\ \

II. {\it{The problem of the neutrino oscillation}}.

The neutrino oscillations, i.e. the mutual oscillating transitions
of the neutrinos of different generations, are observed for a long
time now. The common explanation of the phenomenon is based on the
assumption the neutrino mass matrix is non-diagonal. Moreover, in
order to match all the experimental evidences, the extra neutrino
fields are introduced, which are sterile regarding to all
interactions (naturally, except gravitational one). The sterile
neutrinos cannot be observed directly: they are coupled to the three
known neutrino generations only by means of a common mass matrix,
and this is the way they give a contribution to the neutrino
oscillations. The introduction of sterile neutrinos does not exhaust
all difficulties of the theory: possibly, the most confounding
factor of the theory consists in the fact that the electroweak
interaction becomes nonrenormalizable one.

The detailed description of the neutrino oscillations experiments
and theory can by found, for example, in \cite{12}, \cite{13}, and
in numerous references there.

Now let's consider the possibility of another physics which may provide the neutrino
oscillations. The basis for this physics is the Wilson fermion
doubling phenomenon on irregular lattices discussed above.

Let's consider the scattering of the usual normal long-wavelength
electron neutrino quantum with the momentum $k_e$, $|k_e|\ll l_P^{-1}$, by the condensate
of muon irregular quanta. Suppose the interaction is mediated by the
gravitational field \footnote{The interaction mechanism of the scattering process is unclear, but it seems to me that the interaction is mediated by the
gravitational field}. This scattering process is pictured in
Fig.~\ref{fig:2}. Obviously, time-mean value of the irregular quantum momentum is equal to zero, and the corresponding necessary minimal averaging time $\tau\sim l_P$. This means that the latter has zero momentum in the interaction process of the long-wavelength neutrino quantum with neutrino irregular excitation. Suppose also that vacuum is translation invariant. Then the scattering process in Fig.~\ref{fig:2} conserve the momentum of the long-wavelength
neutrino: $k_{\mu}=k_{e}$. The same process as in Fig.~\ref{fig:2} takes place under $\nu_e$ and $\nu_{\mu}$ interchanging.
Finally, we conclude that the neutrino oscillations should be
observed since there are mutual transitions of the electron and muon
neutrinos with fixed and equal momenta.

\begin{acknowledgments}

This work was supported by SS-3139.2014.2.
\end{acknowledgments}


\begin{thebibliography}{11}
\expandafter\ifx\csname natexlab\endcsname\relax\def\natexlab#1{#1}\fi
\expandafter\ifx\csname bibnamefont\endcsname\relax
  \def\bibnamefont#1{#1}\fi
\expandafter\ifx\csname bibfnamefont\endcsname\relax
  \def\bibfnamefont#1{#1}\fi
\expandafter\ifx\csname citenamefont\endcsname\relax
  \def\citenamefont#1{#1}\fi
\expandafter\ifx\csname url\endcsname\relax
  \def\url#1{\texttt{#1}}\fi
\expandafter\ifx\csname urlprefix\endcsname\relax\def\urlprefix{URL }\fi
\providecommand{\bibinfo}[2]{#2}
\providecommand{\eprint}[2][]{\url{#2}}

\bibitem[{\citenamefont{Wilson}(1975)}]{1}
\bibinfo{author}{\bibfnamefont{K.}~\bibnamefont{Wilson}},
  \bibinfo{journal}{Erice lectures notes} \textbf{\bibinfo{volume}{CLNS-321}}
  (\bibinfo{year}{1975}).


\bibitem[{\citenamefont{Kogut and Susskind}(1975)}]{2}
\bibinfo{author}{\bibfnamefont{J.}~\bibnamefont{Kogut}} \bibnamefont{and}
  \bibinfo{author}{\bibfnamefont{L.}~\bibnamefont{Susskind}},
  \bibinfo{journal}{Physical Review D} \textbf{\bibinfo{volume}{11}},
  \bibinfo{pages}{393} (\bibinfo{year}{1975}).

\bibitem[{\citenamefont{Susskind}(1977)}]{3}
\bibinfo{author}{\bibfnamefont{L.}~\bibnamefont{Susskind}},
  \bibinfo{journal}{Physical Review D} \textbf{\bibinfo{volume}{16}},
  \bibinfo{pages}{3031} (\bibinfo{year}{1977}).

\bibitem[{\citenamefont{Luscher}()}]{4}
\bibinfo{author}{\bibfnamefont{M.}~\bibnamefont{Luscher}},
  \bibinfo{journal}{arXiv:hep-th/0102028}.

\bibitem[{\citenamefont{Nielsen and Ninomiya}(1981{\natexlab{a}})}]{5}
\bibinfo{author}{\bibfnamefont{H.}~\bibnamefont{Nielsen}} \bibnamefont{and}
  \bibinfo{author}{\bibfnamefont{M.}~\bibnamefont{Ninomiya}},
  \bibinfo{journal}{Nuclear Physics B} \textbf{\bibinfo{volume}{185}},
  \bibinfo{pages}{20} (\bibinfo{year}{1981}{\natexlab{a}}).

\bibitem[{\citenamefont{Nielsen and Ninomiya}(1981{\natexlab{b}})}]{6}
\bibinfo{author}{\bibfnamefont{H.}~\bibnamefont{Nielsen}} \bibnamefont{and}
  \bibinfo{author}{\bibfnamefont{M.}~\bibnamefont{Ninomiya}},
  \bibinfo{journal}{Nuclear Physics B} \textbf{\bibinfo{volume}{193}},
  \bibinfo{pages}{173} (\bibinfo{year}{1981}{\natexlab{b}}).

\bibitem[{\citenamefont{Vergeles}(2006)}]{7}
\bibinfo{author}{\bibfnamefont{S.N.}~\bibnamefont{Vergeles}},
  \bibinfo{journal}{Nuclear Physics B} \textbf{\bibinfo{volume}{735}},
  \bibinfo{pages}{172} (\bibinfo{year}{2006}).

\bibitem[{\citenamefont{Vergeles}(2008)}]{8}
\bibinfo{author}{\bibfnamefont{S.N.}~\bibnamefont{Vergeles}},
  \bibinfo{journal}{JETP} \textbf{\bibinfo{volume}{106}}, \bibinfo{pages}{46}
  (\bibinfo{year}{2008}).

\bibitem[{\citenamefont{Vergeles}(1979)}]{9}
\bibinfo{author}{\bibfnamefont{S.N.}~\bibnamefont{Vergeles, unpublished, quoted in:}}
\bibinfo{author}{\bibfnamefont{A.A.}~\bibnamefont{Migdal}},
  \bibinfo{journal}{Physics Letters B} \textbf{\bibinfo{volume}{81}}, \bibinfo{pages}{37}
  (\bibinfo{year}{1979}).

\bibitem[{\citenamefont{Fujikawa}(1979)}]{10}
\bibinfo{author}{\bibfnamefont{K.}~\bibnamefont{Fujikawa}},
  \bibinfo{journal}{Physical Review Letters} \textbf{\bibinfo{volume}{42}}, \bibinfo{pages}{1195}
  (\bibinfo{year}{1979}).

\bibitem[{\citenamefont{Atiyah and Singer}(1984)}]{11}
\bibinfo{author}{\bibfnamefont{M.F.}~\bibnamefont{Atiyah}} \bibnamefont{and}
  \bibinfo{author}{\bibfnamefont{I.M.}~\bibnamefont{Singer}},
  \bibinfo{journal}{Proc.Nat.Acad.Sci.} \textbf{\bibinfo{volume}{81}},
  \bibinfo{pages}{2597} (\bibinfo{year}{1984}).

\bibitem[{\citenamefont{Troitsky}(2012)}]{12}
\bibinfo{author}{\bibfnamefont{S.}~\bibnamefont{Troitsky}},
  \bibinfo{journal}{UFN} \textbf{\bibinfo{volume}{182}}, \bibinfo{pages}{77}
  (\bibinfo{year}{2012}).

\bibitem[{\citenamefont{Kopp et~al.}(2013)\citenamefont{Kopp, Machado, Maltoni,
  and Schwetz}}]{13}
\bibinfo{author}{\bibfnamefont{J.}~\bibnamefont{Kopp}},
  \bibinfo{author}{\bibfnamefont{P.~A.} \bibnamefont{Machado}},
  \bibinfo{author}{\bibfnamefont{M.}~\bibnamefont{Maltoni}}, \bibnamefont{and}
  \bibinfo{author}{\bibfnamefont{T.}~\bibnamefont{Schwetz}},
  \bibinfo{journal}{Journal of High Energy Physics}
  \textbf{\bibinfo{volume}{2013}}, \bibinfo{pages}{050} (\bibinfo{year}{2013}).



\end{thebibliography}

\end{document}